# Multifunctional $L1_0$-Mn$_{1.5}$Ga films with ultrahigh coercivity, giant perpendicular magnetocrystalline anisotropy and large magnetic energy product

Lijun Zhu, Shuaihua Nie, Kangkang Meng, Dong Pan, Jianhua Zhao* and Houzhi Zheng

*State Key Laboratory of Superlattices and Microstructures, Institute of Semiconductors, Chinese Academy of Sciences, P.O.BOX 912, Beijing 100083, China*

*Athors to whom correspondence should be addressed. E-mail: jhzhao@red.semi.ac.cn

Magnetic materials with high coercivity ($H_c$), perpendicular magnetic anisotropy (PMA) and magnetic energy product have great application potential in ultrahigh-density perpendicular magnetic recording media,[1-5] high-performance permanent magnets[6-8] and ferromagnetic electrodes of spintronic devices with high magnetic-noise immunity and thermal stability.[9-12] Moreover, many interesting physical phenomena like efficient spin injection into semiconductors from ferromagnetic contact in remanence, giant anomalous Hall effect, and long-lived ultrafast spin procession have been observed in materials with large PMA.[13-15] Among the various PMA materials, noble-metal-based $L1_0$-ordered FePt, FePd and CoPt granular films may exhibit high $H_c$ and PMA.[1,16-20] However, the difficulty in attaining (001)-orientation, indispensable high-temperature (>500 °C) annealing process to form $L1_0$ ordering, and large surface roughness due to their granular textures greatly limit their applications in both magnetic storage and spintronics.[16-20] On the other hand, rare-earth magnets provide the backbone of many products including computers, mobile phones, electric cars and wind-powered generators.[7] However, because of both the increasingly limited availability and high costs of mining and processing of rare-earth elements, new kind of magnetic materials free of rare-earth elements are still urgently needed to replace the expensive rare-earth magnets widely used today.

In the past two decades, noble-metal-free perpendicular magnetized Mn$_x$Ga (1<$x$<1.8) alloy films with $L1_0$ (so-called δ-phased) structures (**Fig.1**a) have gained increasing attention for possible applications in ultrahigh-density magnetic recording media and permanent magnets and spintronics. $L1_0$-Mn$_{50}$Ga$_{50}$ homogeneous film was theoretically predicted to have $H_c$ of 64 kOe, large PMA of 26 Merg/cc, moderate magnetization of 2.51$\mu_B$/Mn and ultralow Gilbert damping constant of 0.0003.[15,21,22] However, although the growth and characterization of $L1_0$-Mn$_x$Ga films have been studied intensively on GaAs, MgO, Cr-MgO, GaN, GaSb, Si and Al$_2$O$_3$ substrates,[15,23-28] only a few of those films on GaAs, MgO and Cr-MgO demonstrated PMA and relatively small $H_c$ no more than 5 kOe at room temperatures;[15,23,24,28] few of others, unfortunately, showed PMA or high room-temperature $H_c$. The quest for $L1_0$-Mn$_x$Ga films with the fascinating theory-predicted properties remains a major challenge, the solution of which could allow for further advances in ultrahigh-density perpendicular magnetic storage, cost-effective permanent magnets and variety of high magnetic-noise-immune and thermal-stable spintronics devices, such as spin-torque perpendicular magnetic random access memories (MRAMs) and oscillators and high-magnetic-field sensors.[9-12]

In this work we present the first report of the pronounced properties including perpendicular $H_c$ tunable from 8.1 to 42.8 kOe, perpendicular magnetocrystalline anisotropy with a maximum of 22.9 Merg/cc, energy product up to 2.60 MGOe, squareness exceeding 0.94, and magnetization controllable from 27.3 to 270.5 emu/cc observed at 300 K in homogeneous $L1_0$-Mn$_{1.5}$Ga epitaxial films on GaAs (001) (see **Fig. 1**b). These room-temperature magnetic characteristics make such noble-metal-free and rare-earth-free $L1_0$-Mn$_{1.5}$Ga films multifunctional and economical material not only for ultrahigh-density magnetic storage and spintronic devices with high magnetic-noise immunity and thermal stability, but also for possible permanent magnets applications.

As representatively shown in **Fig. 1**c, the high-resolution transmission electron microscopy (TEM) image indicates both a well-textured $L1_0$-Mn$_{1.5}$Ga layer with $c$-axis perpendicular to GaAs substrate and an abrupt interface. Moreover, the epitaxial relationship of MnGa(001)[100]||GaAs(001)[110] can be derived from high-resolution TEM image, in consistent with *in situ* reflection high-energy electron diffraction (RHEED) patterns during growth of Mn$_{1.5}$Ga. The low-magnification TEM image in the inset of **Fig. 1**c reveals both the good homogeneity and the designed thickness value of 48±0.5 nm of Mn$_{1.5}$Ga layer.

**Figure 1**d shows the synchrotron radiation X-ray diffraction (XRD) $\theta$-$2\theta$ patterns of Mn$_{1.5}$Ga films fabricated at different temperatures ($T_s$) from 100 to 300 °C. Besides the (002) and (004) peaks of GaAs substrates, only the superlattice (001) and fundamental (002) peaks of $L1_0$-Mn$_{1.5}$Ga films can be observed in the range of 20° to 70°, indicating that the $Mn_{1.5}Ga$ films prepared below 300 °C are good single-crystalline films with $c$-axis along the normal direction, in agreement with both RHEED patterns and cross-sectional TEM images. However, the films grown beyond 350 °C are not phase-pure since some unknown peaks appear in the XRD patterns (not shown here). The possible reaction between GaAs substrates and Mn atoms at high temperature may account for those peaks irrelevant to $L1_0$ phase.

Both perpendicular and in-plane hysteresis loops of 100 °C-grown $Mn_{1.5}Ga$ film measured at 300 K is typically given in **Fig. 2**a. A nearly square loop with ultrahigh $H_c$ of 42.8 kOe and remnant magnetization ($M_r$) of 27.3 emu/cc is observed from the perpendicular $M$-$H$ curve. However, the in-plane $M$-$H$ curve exhibits almost anhysteretic loop, zero remnant magnetization and high saturation field exceeding 90 kOe. This observation offers a direct evidence for the giant perpendicular anisotropy. The same feature holds for all the films grown at temperatures between 100 and 300 °C. The perpendicular hysteresis loops of $Mn_{1.5}Ga$ films grown at different $T_s$ are shown in **Fig. 2**b. Among them, the very small step (~2.1 emu/cc) around zero magnetic field in $M$-$H$ curve of 100 °C-grown $Mn_{1.5}Ga$ film (also see the blue curve in **Fig. 2**a) is attributed to some magnetically soft domains with poor chemical ordering. No such steps are observable in other $M$-$H$ curves due to improvement of the crystalline quality at high $T_s$. It should be also pointed out that the rounding of some hysteresis loops can be understood by "non-coherent rotation" related to structural imperfection and some degree of inhomogeneity. From **Fig. 2**b, it can be found that perpendicular $H_c$ and $M_s$ are strongly dependent on $T_s$ in contrast to $M_r/M_s$. Details of the dependence of $M_r/M_s$, $M_s$, and $H_c$ on $T_s$ are summarized in **Fig. 2**c-e. From 100 to 300 °C, $M_r/M_s$ keeps a nearly invariantly value of 0.94, very different from that in $L1_0$-$Mn_{1.5}Ga$ films grown on GaN, GaSb, Si and $Al_2O_3$.[25-27] $M_s$ increases monotonically from 27.3 to 270.5 emu/cc as $T_s$ increases, which is much smaller than the calculated value of 2.51 $\mu_B$/Mn (~550 emu/cc) for stoichiometric $L1_0$-MnGa.[21] Both the overall strains induced by the short c axis[22] (see **Fig. 3**d) and off-stoichiometry should be responsible for the relatively small magnetization. As theoretically predicted in ref. 21, excess Mn atoms will align antiparallel to the rest and result in compensation.

As shown in **Fig. 2**e, $H_c$ has a maximum of 42.8 kOe at 100 °C, then drops to 8.1 kOe at 250 °C, and finally climbs up to 10.8 kOe at 300 °C. The highest $H_c$ of 42.8 kOe may support a giant magnetoresistance (GMR) sensor taking the 100 °C-grown $L1_0$-$Mn_{1.5}Ga$ film as the perpendicular ferromagnetic electrode to measure large filed up to 42 kOe in the case proposed in ref. 11. Importantly, the ultrahigh $H_c$ could also make related devices highly immune to external magnetic noises. It is also quite striking that so high coercive fields found in these $L1_0$-$Mn_{1.5}Ga$ epitaxial films, since the reported value of $L1_0$-$Mn_xGa$ films is quite small, no more than 5 kOe.[15,23-28] Interestingly, our $L1_0$-$Mn_{1.5}Ga$ films are of very continuous textures and smooth surfaces, thus, the mechanism of ultrahigh $H_c$ should be quite different from that in FePt, FePd and CoPt granular films[16-20] or $D0_{22}$-$Mn_xGa$ (2<$x$<3) nanostructured films on Si/SiO$_2$,[6,7] in which $H_c$ is linked to magnetic separation of individual nanoparticles.

Uniaxial perpendicular magnetocrystalline anisotropy constant ($K_u$) usually plays a central role in determining $H_c$ of PMA materials. To evaluate the PMA properties quantitatively and investigate the origins of the ultrahigh $H_c$ exhibited in these $L1_0$-$Mn_{1.5}Ga$ films, we estimated $K_u$ from the relation $K_u=K^{eff}+2\pi M_s^2$, where $K^{eff}$ is the effective perpendicular magnetic anisotropy constant deduced from the area enclosed between the magnetization curves in applied fields parallel and perpendicular to the film plane. **Fig. 3**a displays $K_u$ as a function of $T_s$. With increasing $T_s$, $K_u$ goes up monotonically and reaches the maximum value at 250 °C and then decreases with further increasing $T_s$. The maximum $K_u$ of 22.9 Merg/cc is roughly consistent with the theoretical value of 26 Merg/cc for $L1_0$-$Mn_{50}Ga_{50}$,[21] while much larger than the reported value of 15 Merg/cc in $L1_0$-$Mn_{1.54}Ga$ films on MgO,[15,28] 0.89 Merg/cc in $D0_{22}$-$Mn_3Ga$ films,[29,30] and 2.1 Merg/cc in CoFeB ultrathin films,[10] and that of common noble-metal-based $L1_0$-ordered granular films.[17-19] The smallest dimension $d$ for which the thermal stability condition $K_u d^3/k_B T \geq 60$ is satisfied at 300 K when $K_u$ =22.9 Merg/cc is 4.77 nm. In other words, the giant $K_u$ of 22.9 Merg/cc in our $L1_0$-$Mn_{1.5}Ga$ films supports nanoscale spintronic devices, such as spin-torque MRAMs bits, down to 5 nm in size and high-density bit-patterned recording with areal density up to 30 Tb inch$^{-2}$ and 60-year thermal stability.

According to the Stoner-Wolhfarth model,[31] $K_u$ determines the upper limit of $H_c$ in a perfectly homogeneous uniaxial magnet. For our $L1_0$-$Mn_{1.5}Ga$ films, unfortunately, the change tendency of $K_u$ with $T_s$ is completely different from that of $H_c$ versus $T_s$. Thus, it can be considered that there exist other important contributions to $H_c$, though the giant $K_u$ must be responsible for the ultrahigh $H_c$ in common. Just as Brown's paradox[32] revealed, another possible contribution to the unusual $H_c$ should be imperfection of the films including chemical disorder, lattice defects and strains. We, therefore, carefully evaluated the degree of chemical ordering ($S$), the full width at half maximum (FWHM) of $Mn_{1.5}Ga$ (002) peaks and the lattice constant c of these films from the XRD patterns in **Fig. 3**b-d. $S$ standing for the probability of correct site occupation in the crystal structure often has significant influence on the magnetic properties of the $L1_0$-ordered alloy films.[16] The $S$ value is defined by the square root of the intensity ratio of (001) superlattice and (002) fundamental reflections, divided by the theoretical ratio.[16] As shown in **Fig. 3**b, the change tendency of $S$ with $T_s$ is much similar to that of $K_u$ versus $T_s$, while exactly opposite to that of $H_c$ versus $T_s$ in **Fig. 2**e. Therefore, the chemical disorder, probably makes a crucial contribution to coercivity although gives decrease $K_u$ at the same time. $T_s$ dependence of FWHM of $L1_0$-Mn$_{1.5}$Ga (002) peaks is shown in **Fig. 3**c. As $T_s$ increases from 100 to 250 °C, FWHM decreases monotonically, suggesting that the crystalline quality be improved and defects decrease. When $T_s$ further increases, FWHM goes up, indicating a drop of crystalline quality and an increase in defects. Evidently, FWHM has nearly the same $T_s$ dependence with perpendicular $H_c$, indicating probably contribution to the coercivity from the lattice defects. Lattice constant $c$ for Mn$_{1.5}$Ga films plotted as a function of $T_s$ in **Fig. 3**d slightly varies from 3.47 to 3.51 Å, which is much smaller than the bulk value of 3.69 or 3.642 Å,[24,33] indicating the strong tensile strains in these films. Theoretical work by Yang *et al*. found that reducing lattice parameter $c$ leads to a smaller magnetic moment.[22] We infer that the strong overall tensile strains should also account for the ultrahigh $H_c$ in our films.

From all the factors discussed above, we deduce that the ultrahigh $H_c$ in these $L1_0$-Mn$_{1.5}$Ga films originates from the combination of giant perpendicular magnetic anisotropy and imperfection including chemical disorder, lattice defects and overall tensile strains. Importantly, the imperfection contribution to $H_c$ seems to be the same order of magnitude as the contribution from the magnetic anisotropy since $H_c$ can be tailed so obviously over the large scale from 8.1 to 42.8 kOe.

It is also worth mentioning that the noble-metal-free $L1_0$-Mn$_x$Ga alloys have much lower magnetic damping constants than those of all the other PMA materials to our best knowledge due to both the low density of states at the Fermi level and the weak spin-orbit interaction in light 3$d$ elements Mn and Ga, just as indicated both experimentally and theoretically in ref. 15. Therefore, we infer that our $L1_0$-Mn$_{1.5}$Ga films also have very low damping constants, and detail studies by time-resolution magneto-optic Kerr effect measurements are in progress to confirm the very low damping constant and will be given elsewhere.

Finally we want to mention that magnetic materials with high coercivity, linear demagnetization induction curve (*B-H*) in the second quadrate and large magnetic energy product are also promising in permanent magnets applications [8]. From this point of view, we investigated demagnetization curves and confirmed the linear demagnetization induction curves (not shown here), which are directly related to the high squareness of *M-H* curves. Further, the magnetic energy product $(BH)_{max}$ was evaluated in **Fig. 4**a and compared with the ideal value $(2\pi M_s)^2$ in **Fig. 4**b. $(BH)_{max}$ climbs up monotonically from 0.02 to 2.60 MGOe as $T_s$ increases and almost the same as ideal value $(2\pi M_s)^2$ when $T_s$<250 °C because of the very high $M_r/M_s$. $(BH)_{max}$ at $T_s$ = 300 °C is slightly smaller than the ideal value probably due to the rounding at corner of *M-H* loops and the small difference between $M_s$ and $M_r$ ($M_r/M_s$ = 0.94). The largest energy product up to 2.60 MGOe are observed in the film grown at 300 °C, which is larger than that previously reported in bulk polycrystalline D0$_{22}$-Mn$_{2\sim3}$Ga alloys (<5.5 kJ/m$^3$ or 0.68 MGOe)[29] and steels magnets (~1 MGOe)[8], and also comparable with hard ferrite magnets (~3 MGOe)[8], while still smaller than that in the rare-earth based magnets like SmCo$_5$ (22 MGOe)[34] and Nd$_2$Fe$_{14}$B (18 or 56 MGOe)[8,35]. Here, it is the small $M_s$ who limit the magnitude of $(BH)_{max}$ of the films, though relatively small $M_s$ is appreciated in spin-torque applications. Fortunately, it can also be competitive with rare-earth based SmCo$_5$ and Nd$_2$Fe$_{14}$B magnets if we can tune the $M_s$ to beyond 550 emu/cc by adjusting compositions, optimizing growth conditions or post-annealing. Related studies are in progress and will be given elsewhere. Therefore, the rare-earth-free compositions, high coercivity, linear demagnetization induction curve and large energy product together make this kind of $L1_0$-Mn$_x$Ga films promising to be developed for economical permanent magnets.

In summary, we have presented the homogenous noble-metal-free and rare-earth-free $L1_0$-Mn$_{1.5}$Ga epitaxial films on GaAs (001) with perpendicular coercivity tunable from 8.1 to 42.8 kOe, perpendicular magnetocrystalline anisotropy up to 22.9 Merg/cc, magnetic energy product up to 2.60 MGOe, magnetization controllable from 27.3 to 270.5 emu/cc. These room-temperature magnetic characteristics make our $L1_0$-Mn$_{1.5}$Ga films multifunctional as pronounced and cost-effective alternative for not only perpendicular magnetic recording bits with areal density over 30 Tb inch$^{-2}$ and thermal stability over 60 years, but variety of novel devices with high magnetic-noise immunity and thermal stability like spin-torque MRAMs and oscillators pillars below 5 nm in dimension, and GMR sensors able to measure filed up to 42 kOe. Moreover, this kind of material is also promising to be developed for economical permanent magnets.

*Experimental*

**Sample fabrication**: Each $L1_0$-Mn$_x$Ga film was deposited on a 150 nm-GaAs-buffered semi-insulating GaAs (001) substrate at different temperature from 100 to 400 °C by molecular-beam epitaxy, and capped with a 4 nm aluminum layer after cooling down to room temperature to avoid oxidization. The compositions were designed by controlling Mn and Ga fluxes during growth, and later verified by x-ray photoelectron spectroscopy measurements to be Mn$_{60}$Ga$_{40}$ (shorten as Mn$_{1.5}$Ga).

**Characterization:** The crystalline structures were characterized by *in situ* reflection high-energy electron diffraction (RHEED), *ex-situ* cross-sectional high-resolution transmission electron microscopy (TEM) and synchrotron radiation x-ray diffraction (XRD) with wave length of 1.5406 Å. The magnetic properties were characterized by superconducting quantum interference devices (SQUID) up to 5 T and physical property measurement system (PPMS) up to 9 T.

*Acknowledgements*

We thank Dr. Q. Cai at Beijing Synchrotron Radiation Facility and Professor H. H. Zhou at Beijing Electron Microscopy Center for their assistance with XRD and TEM measurements, respectively. This work is supported partly by the National Natural Science Foundation of China under Grant Nos. 60836002 and 10920101071.

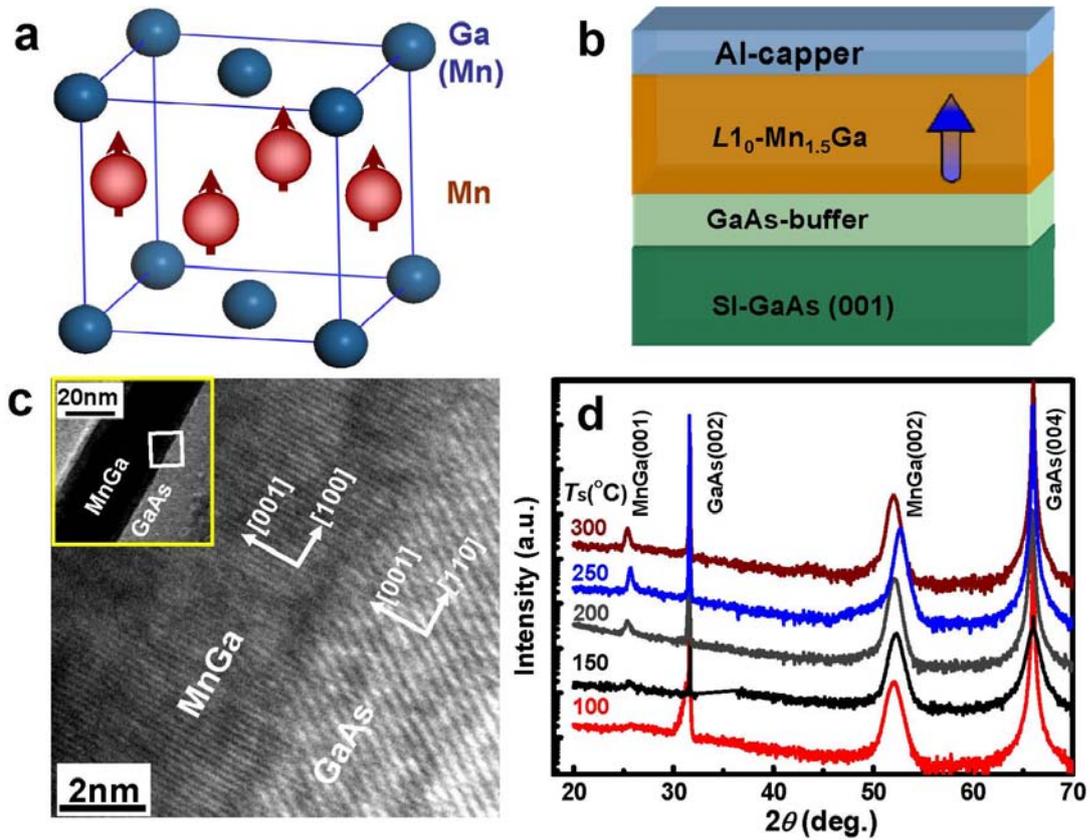

**Figure 1. a)** Unit cell of $L1_0$-$Mn_xGa$ (1<x<1.8). **b)** Schematic diagram of the sample structures, blue arrow stands for the perpendicular magnetization of $Mn_{1.5}Ga$ film. **c)** A high-resolution cross-sectional TEM image at $L1_0$-$Mn_{1.5}Ga$/GaAs interface with epitaxial relationship of $Mn_{1.5}Ga[100](001)/GaAs[110](001)$. Inset: Low-magnification image, white rectangle shows the region of high magnification image. d) Synchrotron radiation X-ray diffraction $\theta$-$2\theta$ patterns of $L1_0$-$Mn_{1.5}Ga$ films grown at 100, 150, 200, 250 and 300 °C, respectively.

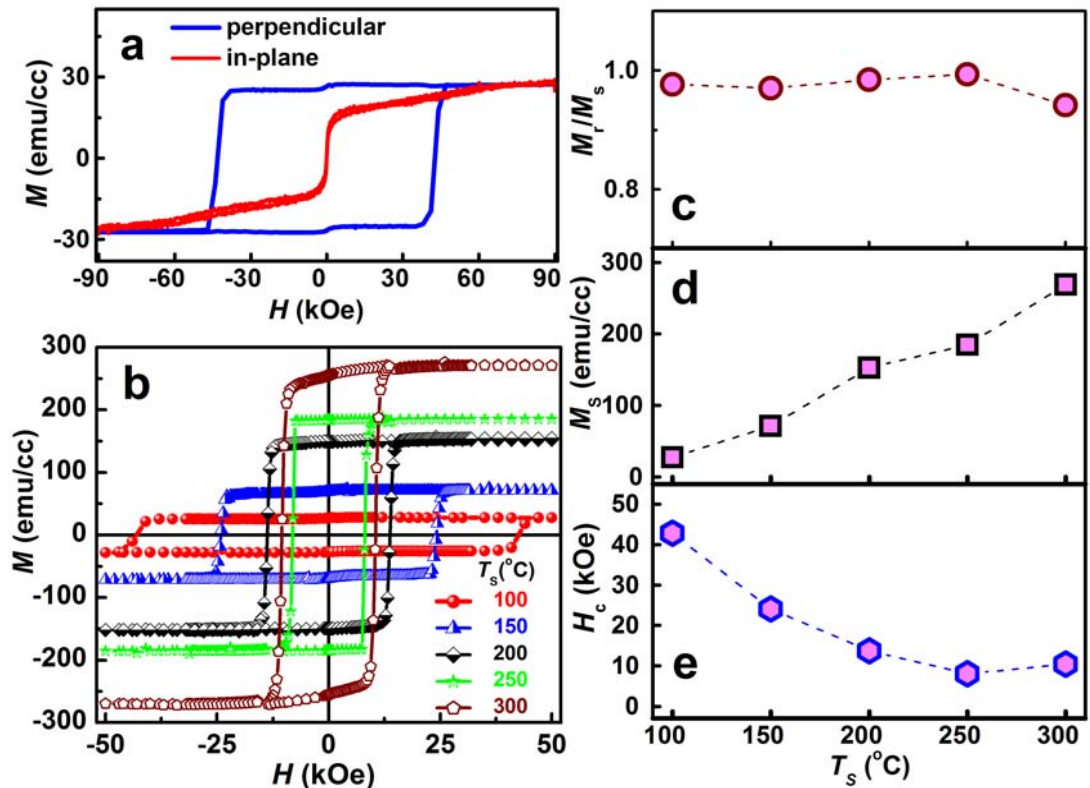

**Figure 2. a)** Perpendicular and in-plane hysteresis loops of the 100℃-grown $L1_0$-$Mn_{1.5}Ga$ film measured by PPMS. **b)** Perpendicular hysteresis loops of the $L1_0$-$Mn_{1.5}Ga$ films grown at different $T_s$ measured by SQUID. **c)** $M_r/M_s$, d) $M_s$ and e) $H_c$ plotted as a function of $T_s$.

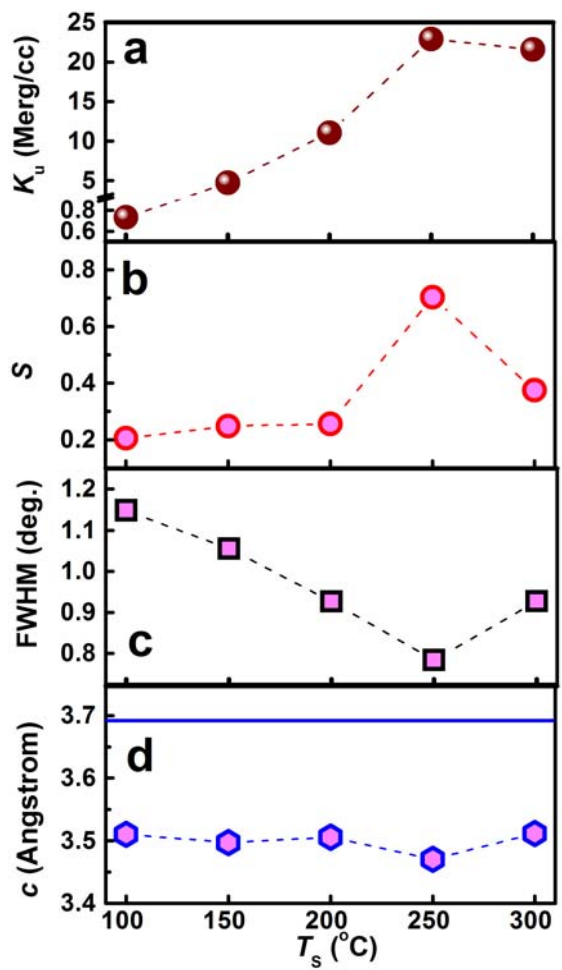

**Figure 3.** $T_s$ dependence of a) $K_u$, b) $S$, c) FWHM of $Mn_{1.5}Ga$ (002) peaks and d) lattice parameters $c$ of $L1_0$-$Mn_{1.5}Ga$ films. The blue solid line in d) presents the $c$ value of 3.69 Å for $L1_0$-$Mn_{1.5}Ga$ bulk.

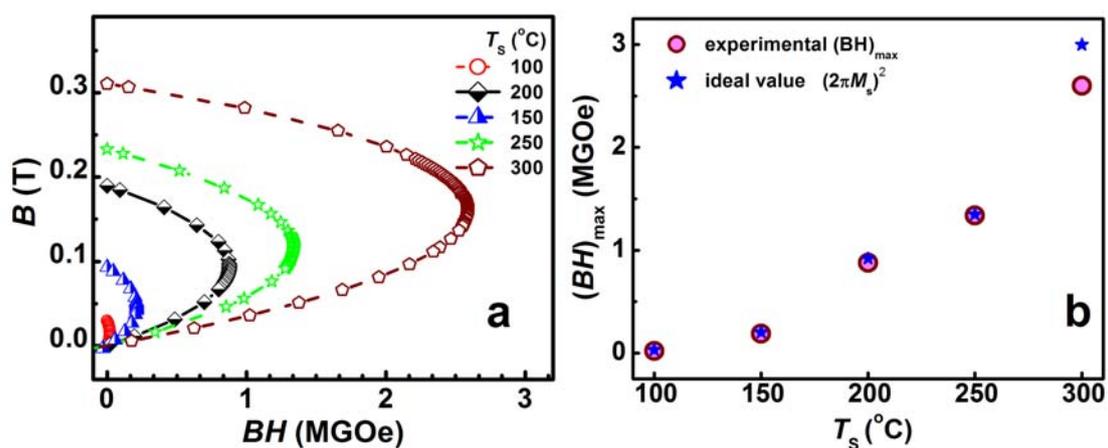

**Figure 4. a)** Magnetic energy products of $L1_0$-$Mn_{1.5}Ga$ films and **b)** comparison of the experimental $(BH)_{max}$ with ideal value $(2\pi M_s)^2$.